\documentclass[preprint, resetfootnote]{aastex701}

\newcommand{\hmi}{{\it SDO}/HMI}
\usepackage{gensymb}
\usepackage{hyperref}
\usepackage{float}
\usepackage{placeins}  
\usepackage{soul}

\begin{document}

\title{Structural Tilting and Depth-Dependent Behavior of Equatorial Rossby Waves}

\correspondingauthor{Oana Vesa} 

\author[orcid=0000-0001-6754-1520,gname=Oana,sname=Vesa]{Oana Vesa}
\affiliation{W.~W.~Hansen Experimental Physics Laboratory, Stanford University, Stanford, CA 94305-4085, USA}
\email[show]{ovesa@stanford.edu}  

\author[0000-0002-6308-872X]{Junwei Zhao}
\affiliation{W.~W.~Hansen Experimental Physics Laboratory, Stanford University, Stanford, CA 94305-4085, USA}
\email{junwei@sun.stanford.edu}

\author[0000-0002-2632-130X]{Ruizhu Chen}
\affiliation{W.~W.~Hansen Experimental Physics Laboratory, Stanford University, Stanford, CA 94305-4085, USA}
\email{rzchen@stanford.edu}

\begin{abstract}
Over the past decade, solar equatorial Rossby waves have been unambiguously identified and are considered potential diagnostics of solar interior dynamics. 
We investigate their inclined structure and temporal evolution in the solar interior across multiple depths using approximately 14.5\,yr of ring-diagram (RD) and time-distance (TD) helioseismology data from {\it SDO}/HMI. 
Normalized phase differences and cross power are computed from filtered spherical harmonic coefficients of radial vorticity to probe the structural tilt and power of Rossby waves.
We find a systematic and robust depth-dependent phase behavior that shows no clear significant correlation with the solar cycle, while the depth-dependent cross power exhibits a positive correlation with the solar cycle for both datasets.
Our results show that deeper depths lead in phase over shallower ones, with increasing negative phases with depth.
We infer that Rossby waves exhibit a retrograde tilt relative to the Sun's rotation that is stable throughout the solar cycle.
Analogous small tilts have been noted in planetary atmospheres and in magnetohydrodynamic simulations of the Sun, indicating that this behavior is not uncommon in rotating, stratified bodies and has implications for angular momentum and energy transport in the solar interior.
\end{abstract}


\section{Introduction}  \label{sec:introduction}

The Sun exhibits a rich spectrum of oscillations across various spatiotemporal scales from the interior to the outer atmosphere.
Among these are Rossby waves, large-scale vorticity oscillations that naturally arise in rotating fluid bodies due to the variation of the Coriolis parameter with latitude \citep{1735_Hadley, 1939_Rossby}.
Seminal works established the existence of Rossby modes in stellar atmospheres \citep[e.g.,][]{1978_Papaloizou_Pringle, 1982_Saio, 1986_Wolff_Blizard}, including that of our Sun.
In the solar context, the generation and excitation of Rossby waves is understood as a natural consequence of the Sun's differential rotation and convective dynamics \citep[e.g.,][]{1986_Wolff_Blizard, 2022_Dikpati_Gilman_Guerrero_etal, 2023_Philidet_Gizon}.
Rossby waves are a subclass of inertial modes that propagate retrograde to the Sun's rotation and that have recently attracted significant attention as probes for solar interior dynamics and solar cycle variability \citep[e.g.,][]{2018_Loptien_Gizon_Birch_etal, 2020_Goddard_Birch_Fournier_Gizon, 2020_Gizon_Fournier_Albekioni, 2023_Waidele_Zhao}.

Equatorial Rossby waves have been identified using a range of helioseismic techniques applied to long-term observations from multiple instruments, including the Helioseismic and Magnetic Imager \citep[HMI;][]{2012_Scherrer_Schou_Bush_etal, 2012_Schou_Scherrer_Bush_etal} onboard the {\it Solar Dynamics Observatory} \citep[{\it SDO};][]{2012_Pesnell_Thompson_Chamberlin}, the Michelson Doppler Imager (MDI) onboard the {\it Solar and Heliospheric Observatory} \citep[{\it SOHO};][]{1995_Scherrer_Bogart_Bush_etal}, and the Global Oscillation Group Network \citep[GONG++;][]{2003_Corbard_Toner_Hill_etal}.
These observations confirmed the unambiguous presence of equatorial Rossby waves with azimuthal orders $3\,\leq\,m\,\leq\,20$ near the solar surface through multiple techniques applied to the radial component of the flow vorticity, including local correlation tracking \citep[e.g.,][]{2018_Loptien_Gizon_Birch_etal, 2021_Hathaway_Upton}, ring-diagram analysis \citep[e.g.,][]{2020_Proxauf_Gizon_Loptien_etal, 2020_Hanson_Gizon_Liang}, time-distance helioseismology \citep[e.g.,][]{2019_Liang-Gizon_Birch_Duvall, 2023_Waidele_Zhao}, and normal-mode coupling \citep[e.g.,][]{2020_Mandal_Hanasoge, 2021_Mandal_Hanasoge_Gizon}.
The observed behavior of these modes agrees with theoretical expectations for classical sectoral Rossby waves \citep{1982_Saio, 1998_Wolff}.

While the detection of equatorial Rossby waves is now well established, their subsurface structure, depth-dependence, and potential radial propagation remain poorly constrained. 
Near-surface characteristics are relatively well-determined; however, attempts to probe deeper subsurface layers have produced divergent results depending on the helioseismic technique employed.
For example, \added{\citet{2021_Hathaway_Upton} reported depth-invariant power amplitudes down to depths of 37\,Mm.
In contrast, \citet{2020_Proxauf_Gizon_Loptien_etal} reported a monotonic decrease of $10\,\%$ with depth down to 8\,Mm, while \citet{2018_Loptien_Gizon_Birch_etal} also found a decrease to 11\,Mm.}
\citet{2024_Mandal_Hanasoge}, on the other hand, observed amplitudes systematically increasing with depth down to \added{$\approx$\,56\,Mm} before decreasing further below.
Limited observations and inconsistent findings leave the radial structure of Rossby waves and their potential for energy transport an open question.

Constraining these depth-dependent properties is particularly important, as magnetic fields and solar cycle variations further modulate Rossby wave properties and dynamics \citep[for a more comprehensive review, see][and references therein]{2021_Zaqarashvili_Albekioni_Ballester_etal}. 
Rossby waves across various solar layers show systematic modulation by solar activity as seen in both numerical simulations and observations.
Theoretical models of Rossby waves within the tachocline using shallow-water magnetohydrodynamics (MHD) frameworks demonstrate that the presence and strength of toroidal magnetic fields alter their propagation properties, generating both fast and slow magneto-Rossby waves that are sensitive to stratification and field strength \citep{2007_Zaqarashvili_Oliver_Ballester_Shergelashvili, 2009_Zaqarashvili_Oliver_Ballester, 2020_Dikpati_Gilman_Chatterjee_etal}.
Observational studies show that equatorial Rossby wave power correlates positively and wave frequencies correlate negatively (i.e., larger retrograde propagation) with heightened solar activity \citep{2023_Waidele_Zhao, 2024_Lekshmi_Gizon_Jain_etal}.
Additionally, the deep interior magneto-Rossby waves have been associated with quasi-periodic variations in solar activity, suggesting that coupling between Rossby waves, differential rotation, and magnetic fields can influence solar cycle dynamics \citep{2010_Zaqarashvili_Carbonell_Oliver_Ballester, 2017_Dikpati_Cally_McIntosh_Heifetz, 2018_Dikpati_McIntosh_Bothun_etal}.

In this work, we focus on the phase information derived from the spherical harmonic coefficients of radial vorticity to probe the depth-dependent structure of equatorial Rossby waves.
By analyzing the cross power and phase differences between subsurface layers derived from helioseismology techniques, we examine how the structural tilt of Rossby waves varies with depth and throughout the solar cycle. 
This work is structured as follows.
The data analysis procedure is described in Sect.\,\ref{sec:data_analysis}.
The results are presented in Sect.\,\ref{sec:results}. 
Lastly, discussion and conclusions follow in Sect.\,\ref{sec:discussion_and_conclusions}.

\section{Data Analysis} \label{sec:data_analysis}

\subsection{Helioseimology Flow Fields} \label{subsec:helioseismology_data_prep}
We use horizontal flows of $v_x(\theta, \phi)$ and $v_y(\theta, \phi)$, where $\theta$ is the co-latitude and $\phi$ is longitude, from the \hmi\ ring-diagram \citep[RD;][]{2011_Bogart_Baldner_Basu_etal_P1, 2011_Bogart_Baldner_Basu_etal_P2} and time-distance helioseismology \citep[TD;][]{2012_Zhao_TD} pipelines.

The RD data \added{span nominal target depths of 1.4\,Mm--16.0\,Mm, in increments of $\approx$\,2.1\,Mm, covering Carrington rotations (CR) 2097--2293, or roughly the period from 2010 May 19 to 2025 January 6.}
These horizontal subsurface flows are derived from $15\,\degree \times 15\,\degree$ tiles (spatial sampling of $7{\fdg}5$\,pixel$^{-1}$) and have a temporal cadence of approximately $27$\,hr (or $\approx$\,1/24 of a synodic rotation).
For each depth, the data is prepared and \added{rearranged to a reference frame rotating at the synodic Carrington frequency of 424.3\,nHz}.

The TD data typically covers the depth from the surface to approximately 20\,Mm with irregular depth intervals, spanning a similar period as the RD data. 
The TD data covers CR 2096--2293, or 2010 May 1 to 2024 December 31.
In this work, we use data corresponding to depths of 0--1\,Mm, 1--3\,Mm, 3--5\,Mm, 5--7\,Mm, 7--10\,Mm, and 10--13\,Mm.
For simplicity, we refer to these depth ranges by their midpoints (e.g., 0.5\,Mm, 2\,Mm, 4\,Mm, 6\,Mm, 8.5\,Mm, and 11.5\,Mm).
These data have a temporal cadence of 8\,hr with a spatial sampling rate of $0\fdg12$ pixel$^{-1}$, substantially higher than the RD sampling rate.
To better match the RD flow maps, hence the follow-up analysis procedure, we average the TD flow fields within $15\degr\times15\degr$ boxes to one flow vector, and average three consecutive flow fields to reduce the temporal cadence to 24\,hr.  
We also choose our final spatial sampling rate to be $7\fdg5$\,pixel$^{-1}$, the same as the RD sampling rate.

The RD and TD radial vorticity $\zeta(\theta, \phi, t)$ data products are computed for each depth following \citet{2023_Waidele_Zhao}, derived directly from the horizontal velocity components.
As Rossby waves obtain maximum amplitudes around the Sun's equator, we restrict ourselves to the equatorial band around $\pm\,22.5\degree$.

\subsection{Radial Vorticity Decomposition and Phase Analysis} \label{subsec:methods}

To investigate the longitudinal depth-dependent structure of equatorial Rossby waves, we perform a spherical harmonic decomposition (SHD) on $\zeta(\theta, \phi, t)$ for each depth $d$, which yields the complex coefficient $\bar{\zeta}_{\ell,m}(t, d)$,
\begin{equation}\label{eqn:spherical_harmonic_decomposition_eqn}
    \bar{\zeta}_{\ell,m}(t, d) = \sum_{\theta, \phi} \zeta_{\ell,m}(\theta, \phi, t, d)\mathcal{Y}_{\ell,m}^*(\theta, \phi)\sin\theta,
\end{equation}
in which $\mathcal{Y}_{\ell,m}^*(\theta, \phi)$ is the complex conjugate of the spherical harmonic function of angular degree $\ell$ and azimuthal order $m$.
In the following analysis, we restrict ourselves to the sectoral modes, i.e., $\ell = m$ \citep[e.g.,][]{1998_Wolff, 2000_Yoshida_Lee}.
For simplicity, we will drop an $m$ in the subscript notation from hereafter.
We note that $\mathcal{Y}_{m}^*(\theta, \phi)$ introduces a phase
\begin{equation}
    \mathcal{Y}_{m}^*(\theta, \phi) \propto e^{-im\phi},
\end{equation}
where $e^{-im\phi}$ describes the azimuthal variation (in longitude) of $m$.
Therefore, $\bar{\zeta}_{m}(t, d)$ carries a time-dependent phase and amplitude that reflects the longitudinal structure of each $m$ for a particular depth $d$.

By analyzing the complex SHD coefficients, we can examine the joint amplitudes and tilt of Rossby waves between the near-surface reference depth and subsurface depths.
To start, we compute the phase $\varphi_{m}(t, d)$ of each SHD coefficient \citep{2005_Knaack_Stenflo} for a particular depth as 
\begin{equation}\label{eqn:phase_eqn}
    \varphi_{m}(t, d) = \arctan\left[ \frac{\Im[\bar{\zeta}_{m}(t,d)]}{\Re[\bar{\zeta}_{m}(t,d)]}\right].
\end{equation}
The phase difference $\Delta\varphi_{m}(t)$ between a reference depth $d_0$ and successive depths $d_i$ is then 
\begin{equation}\label{eqn:phase_difference_eqn}
\Delta\varphi_{m}(t, d_{0i}) = \varphi_{m}(t, d_0) - \varphi_{m}(t, d_i).
\end{equation}
We define a negative phase difference as deeper layers leading shallower ones, corresponding to a \added{retrograde tilt} relative to the Sun's rotation.

Additionally, the cross power magnitude between depths, which is defined as
\begin{equation} \label{eqn:cross_power_magnitude_eqn}
    \mathcal{S}_m(t, d_{0i}) = |\bar{\zeta}_{m}(t,d_0)\bar{\zeta}^*_{m}(t, d_i)|,
\end{equation}
quantifies the joint amplitude of $m$ and the extent to which its spatial structure is shared across depths.
The autocorrelation of the signal with itself for a given depth defines the power $\mathcal{P}_m(t, d)$.
For completeness, the coherence is defined as
\begin{equation} \label{eqn:coherence}
\mathcal{C}_m(t, d_{0i}) = \frac{|\mathcal{S}_m(t, d_{0i})|^2}{\mathcal{P}_m(t, d_0)\mathcal{P}_m(t, d_i)}
\end{equation}
and measures the degree of correlation between depths, ranging from 0 (no coherence) to 1 (perfect coherence).

\subsection{Fourier Space Filtering of Rossby Waves} \label{subsec:filter}
\begin{figure*}[ht!]
    \centering
    \includegraphics[scale=0.5]{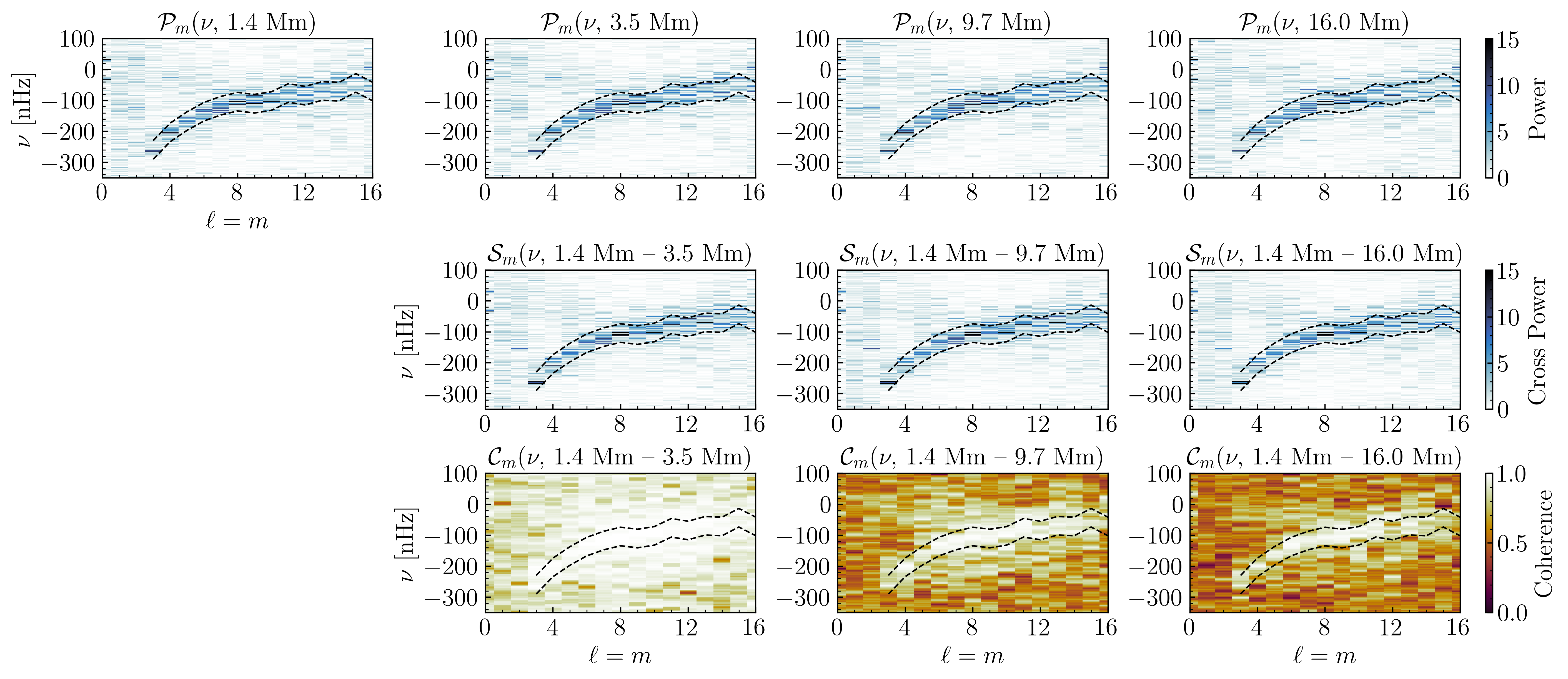}
    \caption{Sectoral power $\mathcal{P}_m(\nu, d)$, cross power $\mathcal{S}_m(\nu, d_{0i})$, and coherence $\mathcal{C}_{m}(\nu, d_{0i})$ spectra of the decomposed RD radial vorticity, focusing on azimuthal orders $3\,\leq\,m\,\leq\,16$. The tracking frequency is $\Omega/(2\pi) = 424.3$\,nHz. Top row: $\mathcal{P}_m(\nu, d)$ for multiple depths from \added{{1.4\,Mm to 16.0\,Mm}}. Middle row: $\mathcal{S}_m(\nu, d_{0i})$ between \added{{1.4\,Mm}} and successive depths indicated in the top row. Bottom row: $\mathcal{C}_{m}(\nu, d_{0i})$ between \added{{1.4\,Mm}} and successive depths. The Rossby wave ridge is indicated by the black dashed lines drawn $\pm$\,30\,nHz around the peak frequencies for each $m$.}
\label{fig:RD_Fourier_Maps}
\end{figure*}

To mitigate the impact of spatial leakage in the quantities derived from the SHD coefficients \citep[e.g.,][]{1998_Hill_Howe}, we first examine the temporal Fourier power $\mathcal{P}_m(\nu, d)$, cross power $\mathcal{S}_m(\nu, d_{0i})$, and coherence $\mathcal{C}_m(\nu, d_{0i})$ as a function of frequency $\nu$ and $m$ for multiple depths.
Following \citet{2020_Proxauf_Gizon_Loptien_etal} and \citet{2023_Waidele_Zhao}, we normalize the Fourier spectra by dividing over $\nu \in [-300,50]$\,nHz for each $m$.
The results for the RD data are presented in Fig.\,\ref{fig:RD_Fourier_Maps} while the qualitatively comparable TD results are shown in Fig.\,\ref{fig:TD_Fourier_Maps}.
The temporal Fourier power spectra serve as a standard reference for identifying Rossby waves.
The observed Rossby wave power is seen at the expected spatial and temporal frequencies for all depths and is consistent with previous findings \citep[e.g.,][]{2018_Loptien_Gizon_Birch_etal, 2024_Mandal_Hanasoge}.

While the normalized power and cross power spectra remain relatively consistent across depths, the coherence exhibits significant variation.
With increasing depth, the coherence becomes distinctly enhanced along the expected Rossby wave ridge while becoming weak and incoherent in the background.
The signal is particularly strong for $6\,\leq\,m\,\leq\,14$, indicating temporal coherence across depths.

To isolate this signal and reduce contamination from neighboring frequencies, we apply a filter in Fourier space around the nominal Rossby wave ridge.
The filter is constructed by identifying the peak frequencies for each $m$, following the weighted average method in \citet{2023_Waidele_Zhao}, and selecting a $\pm$\,30\,nHz window around each peak.
This window, indicated by the black dashed lines in Fig.\,\ref{fig:RD_Fourier_Maps}, encompasses the coherent Rossby wave ridge.
The filter is smoothly tapered at the edges to avoid artificial discontinuities, and \added{we retain the additional Rossby wave ridges arising from spectral leakage to capture the full signal.}
After filtering in Fourier space, we invert back to the time domain of the SHD coefficients to compute the cross power magnitude and phase differences in Sect.\,\ref{subsec:methods}.

\section{Results} \label{sec:results}
\begin{figure*}[ht!]
    \centering
        \includegraphics[scale=0.5]{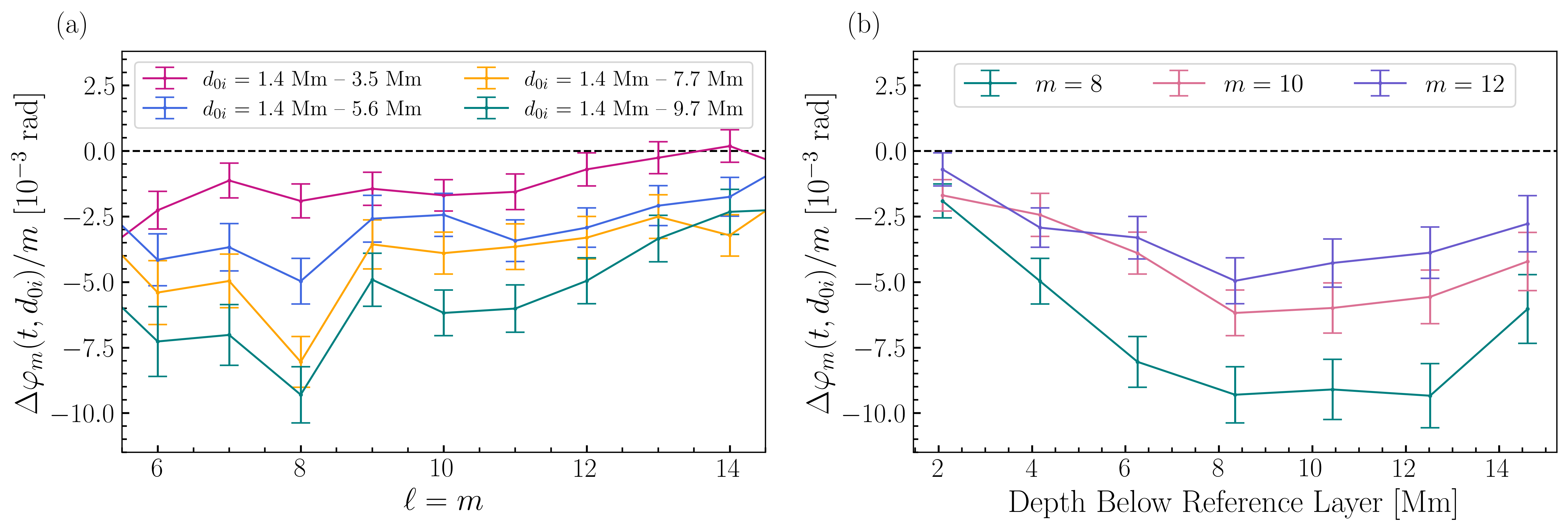}
  \caption{Normalized phase differences $\Delta\varphi_m(t, d_{0i})/m$ derived from the filtered complex coefficients of the decomposed RD radial vorticity between the near-surface reference depth (\added{{1.4\,Mm}}) and several subsurface depths, averaged over approximately 14.5\,yr. (a): $\Delta\varphi_m(t, d_{0i})/m$ versus $\ell=m$, focusing on $6\,\leq\,m\,\leq\,14$. (b): $\Delta\varphi_m(t, d_{0i})/m$ versus depth below the near-surface reference depth for $m$ = $8$, $10$, and $12$. Error bars represent the standard error across time for $\Delta\varphi_m(t, d_{0i})/m$ at each depth and $m$.}\label{fig:RD_longtermphase}
\end{figure*}

\subsection{Long-term Average Behavior of the Depth-Dependent Phases} 
\label{subsec:longterm_phase_average}

To assess the general structural tilt of Rossby waves, we compute the normalized phase differences $\Delta\varphi_m(t, d_{0i})/m$ from the filtered complex SHD coefficients for each $m$ and depth and then average over time.
By computing $\Delta\varphi_m(t, d_{0i})/m$ from the complex SHD coefficients and applying the filtering described in Sect.\,\ref{subsec:filter}, we both restrict the analysis to the coherent signal within the Rossby wave ridge (see Fig.\,\ref{fig:RD_Fourier_Maps}) and mitigate spurious phase contributions from noncoherent background signals and Fourier-induced temporal variability, providing a direct analysis of the tilted wave structure.

Figure\,\ref{fig:RD_longtermphase} shows the average behavior of $\Delta\varphi_m(t, d_{0i})/m$ derived from the RD data across multiple depths, averaged over approximately 14.5\,yr.
The comparable TD data is presented in Fig.\,\ref{fig:TD_longtermphase}.
The error bars represent standard error measurements for $\Delta\varphi_m(t, d_{0i})/m$ at each depth and $m$ \added{{and may be underestimated due to the applied frequency filter.}}

Figure\,\ref{fig:RD_longtermphase}(a) and Fig.\,\ref{fig:TD_longtermphase}(a) show $\Delta\varphi_m(t, d_{0i})/m$ versus $\ell = m$ for the near-surface reference depth and several subsurface depths.
We focus on $6\,\leq\,m\,\leq\,14$, which have high coherence in Fourier space (see Fig.\,\ref{fig:RD_Fourier_Maps}).
In the RD data, the values of $\Delta\varphi_m(t, d_{0i})/m$ are already negative near the surface and become increasingly negative with depth.
In contrast, the TD data show slightly positive $\Delta\varphi_m(t, d_{0i})/m$ between the shallowest depths (0.5\,Mm\,--\,2\,Mm), which then transition to increasingly larger negative phases with depth.
Despite the differences among the shallowest depths, both datasets exhibit qualitatively comparable results in terms of the overall trend and magnitude.

Figure\,\ref{fig:RD_longtermphase}(b) and Fig.\,\ref{fig:TD_longtermphase}(b) show $\Delta\varphi_m(t, d_{0i})/m$ versus distance from the near-surface reference depth for $m =$ 8, 10, and 12.
In both datasets, the general trend of increasingly negative $\Delta\varphi_m(t, d_{0i})/m$ with depth persists until \added{roughly around 8\,Mm,} after which the trend appears to flatten and reverse.
\added{We note, however, that depths greater than 8\,Mm for the RD data are subject to systematic errors \citep[e.g.,][]{2020_Proxauf_Gizon_Loptien_etal}, and the behavior at those depths should be interpreted with caution.}
While all sectoral modes share this qualitative behavior, the magnitude of $\Delta\varphi_m(t, d_{0i})/m$ varies by $m$, suggesting a possible mode-dependent vertical structure.

Collectively, both RD and TD datasets spanning 14.5\,yr show consistently negative $\Delta\varphi_m(t, d_{0i})/m$ with depth, indicating that deeper layers lead shallower ones in phase.
This depth-dependent phase lead suggests that Rossby waves exhibit a retrograde tilt relative to the Sun's rotation, with the magnitude of $\Delta\varphi_m(t, d_{0i})/m$ increasing with depth. 

\subsection{Temporal Evolution of the Depth-Dependent Cross Power and Phases} \label{subsec:temporal_evolution_m}

\begin{figure*}[htb!]
    \centering
    \includegraphics[scale=0.5]{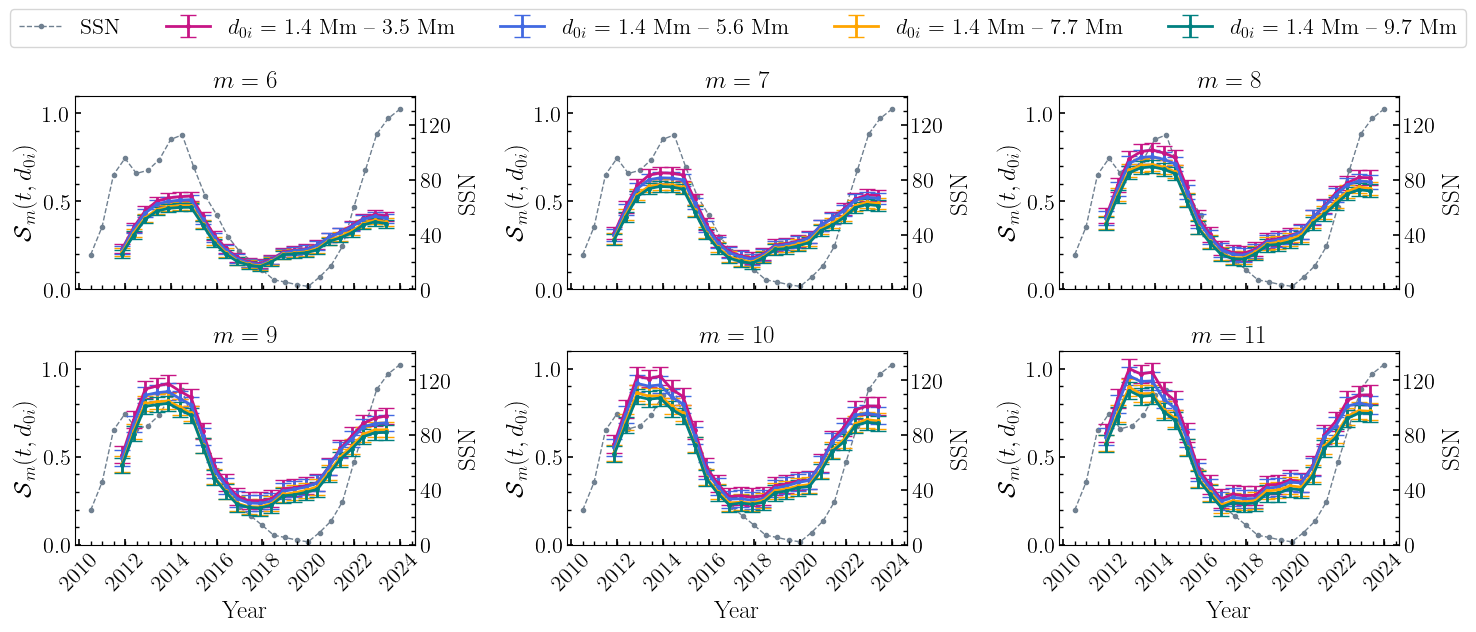}
\caption{Temporal evolution of the normalized cross power $\mathcal{S}_m(t, d_{0i})$ for $6\,\leq\,m\,\leq\,11$ between the near-surface reference depth (\added{{1.4\,Mm}}) and select subsurface depths for the RD data. For all depths, the dataset is segmented into 3\,yr moving windows with a 0.5\,yr step size. Error bars represent standard errors across time segments. The SSN (gray) is overplotted for comparison with the solar cycle.}
    \label{fig:cross_power_temporal}
\end{figure*}

To investigate whether the depth-dependent structure of Rossby waves varies with the solar cycle, we analyze the temporal evolution of the cross power and normalized phase differences for both the RD and TD data.
We divide both datasets into 3\,yr moving windows with a 0.5\,yr step size, allowing us to track temporal changes in $m$ for all depths.
For comparison with the solar cycle, we use the 13-month smoothed sunspot number (SSN)\footnote{\url{https://www.sidc.be/SILSO/infosnmstot}}, which is treated in the same manner as the data.
Our datasets span most of Solar Cycle 24 (SC24) and the rising phase of Solar Cycle 25 (SC25).

\subsubsection{Temporal Evolution and Depth Variation of the Cross Power} \label{subsubsub:temporal_evolution_crosspower}

Figure\,\ref{fig:cross_power_temporal} shows the temporal evolution of the normalized cross power $\mathcal{S}_m(t, d_{0i})$ between the near-surface reference depth and subsequent depths for $6\,\leq\,m\,\leq\,11$ using the RD data.
The TD data is shown in Fig.\ref{fig:TD_cross_power_temporal}.
\added{{The cross power is globally min-max normalized across all time segments, depths, and azimuthal orders $3\,\leq\,m\,\leq\,18$.}}
Error bars represent standard errors across time segments, and the 13-month SSN is overplotted in gray to track the solar cycle.
We focus on azimuthal orders $6\,\leq\,m\,\leq\,11$ because they consistently exhibit the strongest $\mathcal{S}_m(t, d_{0i})$, whereas other $m$ remain comparatively weak.

Analysis of the RD and TD data shows two consistent trends.
First, we find that $\mathcal{S}_m(t, d_{0i})$ decreases with depth for all $m$, though the rate of decrease varies.
On average, $\mathcal{S}_m(t, d_{0i})$ decreases approximately 6$\%$ (TD data) to 12.5$\%$ (RD data) from the reference depth to 8.5\,Mm or \added{9.7\,Mm}, respectively.
Second, the evolution of $\mathcal{S}_m(t, d_{0i})$ is positively correlated with the solar cycle across all depths and $m$. 
We find that the peak in joint amplitudes occurs around solar maximum and declines toward solar minimum of SC24.
A renewed increase in amplitudes is also visible during the rising phase of SC25 in both datasets, further corroborating this positive correlation with the solar cycle.

\subsubsection{Temporal Evolution and Depth Variation of the Phases} \label{subsubsec:temporal_evolution_phase}

\begin{figure*}[ht!]
    \centering
     \includegraphics[scale=0.5]{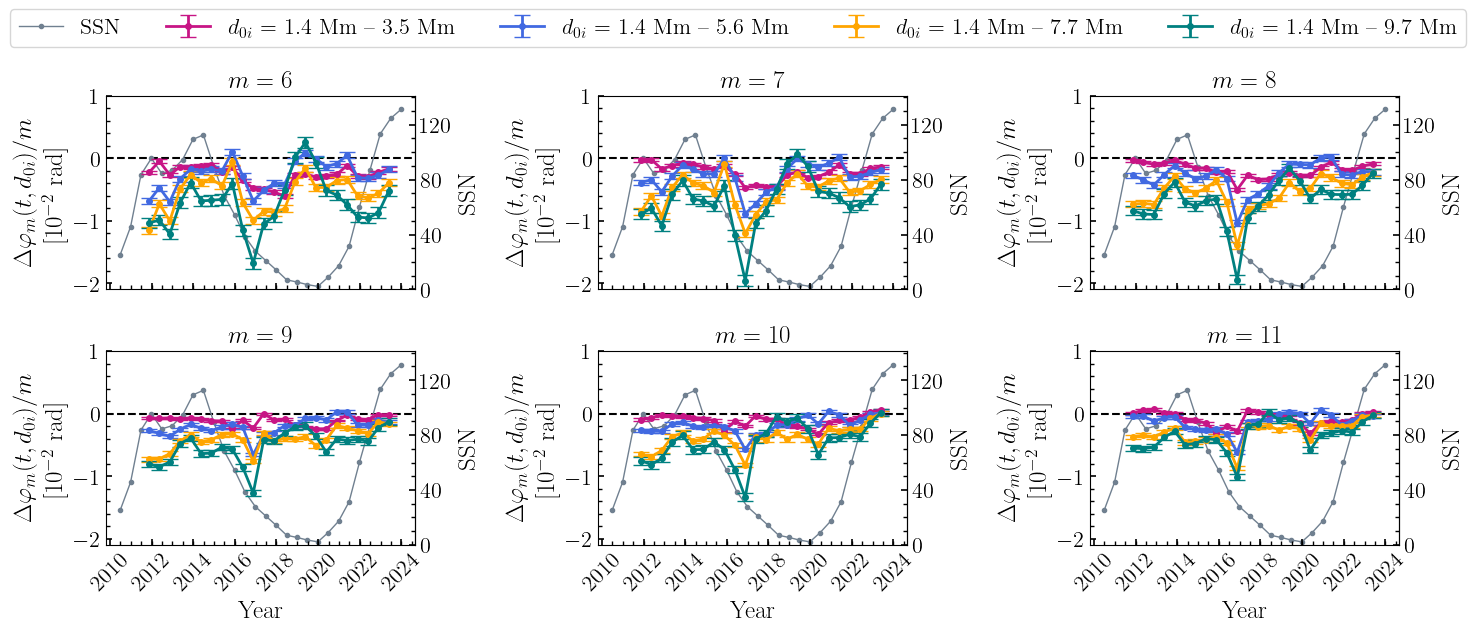}
    \caption{Temporal evolution of the normalized phase differences $\Delta\varphi_m(t, d_{0i})/m$ for $6\,\leq\,m\,\leq\,11$ between the near-surface reference depth (\added{{1.4\,Mm}}) and select subsurface depths for the RD data. For all depths, the dataset is segmented into 3\,yr moving windows with a 0.5\,yr step size. Error bars represent standard errors across time segments. The SSN (gray) is overplotted for comparison with the solar cycle.}
    \label{fig:phase_difference_m}
\end{figure*}

Figure\,\ref{fig:phase_difference_m} shows the temporal evolution of $\Delta\varphi_m(t, d_{0i})/m$ between the near-surface reference depth and subsurface depths for the RD data.
The corresponding TD results are shown in Fig.\,\ref{fig:TD_phase_difference_m}.
For both datasets, error bars represent standard errors computed across time segments, and the 13-month SSN (gray) is shown for reference.
\added{{The error bars in Fig.\,\ref{fig:phase_difference_m} are smaller, reflecting the minimal temporal variation of $\Delta\varphi_m(t, d_{0i})/m$ within individual time segments in contrast to the time-integrated phase differences in Fig.\,\ref{fig:RD_longtermphase}, which exhibit larger error bars than expected from measurement noise alone.}}
For comparison with the most coherent $m$ signals shown in Fig.\,\ref{fig:cross_power_temporal}, we limit our focus to $6\,\leq\,m\,\leq\,11$.

Overall, we find that both RD and TD results exhibit a qualitatively consistent trend of negative $\Delta\varphi_m(t, d_{0i})/m$ for all select $m$ with a clear depth-dependent behavior.
Although $\Delta\varphi_m(t, d_{0i})/m$ remain relatively small in magnitude ($\pm\,0.02$\,rad), these phases are predominantly negative, particularly when the cross power is the strongest (Fig.\,\ref{fig:cross_power_temporal}).
Additionally, $\Delta\varphi_m(t, d_{0i})/m$ systematically shows increasing negative phases with depth, consistent with a clear retrograde tilt of the Rossby wave structure.

Unlike the temporal evolution of the cross power, we observe inconsistent evidence for any strong modulation of $\Delta\varphi_m(t, d_{0i})/m$ by the solar cycle.
In the RD data, we observe weak modulation of $\Delta\varphi_m(t, d_{0i})/m$ by the solar cycle.
While there is a robust depth-dependent behavior, the $\Delta\varphi_m(t, d_{0i})/m$ remain relatively flat across time, except for a pronounced negative spike that occurs around 2017 that coincides with the declining phase of SC24.
This feature, while weaker, is also discernible in the TD data.
In contrast, the TD data exhibit a modest anti-correlation of $\Delta\varphi_m(t, d_{0i})/m$ with the solar cycle, with larger negative phases observed during periods of high solar activity.
Overall, both datasets yield consistent negative phases and depth-dependent trends, suggesting a stable inclined Rossby wave structure.

\section{Discussion and Conclusions} \label{sec:discussion_and_conclusions}

We analyzed the depth-dependent inclined structure of equatorial Rossby waves between the latitudinal band $\pm\,22.5\degree$ and its temporal evolution.
We examined the normalized phase differences and cross power derived from the filtered coefficients of the decomposed radial vorticity from two independent helioseismic datasets (RD and TD), spanning comparable durations and spatial coverage.
Our analysis reveals a robust depth-dependent phase behavior and a solar cycle correlation with the depth-dependent cross power, which is evident in both datasets.

We will first discuss the normalized cross power results, highlighting two key results.
First, the normalized cross power in Fig.\,\ref{fig:cross_power_temporal} and Fig.\,\ref{fig:TD_cross_power_temporal} shows a positive correlation with the solar cycle.
We observe larger joint power during periods of high solar activity, corresponding to solar maximum of SC24 and SC25.
This result is consistent with the time-dependent Fourier power averaged over all $m$ reported by \citet{2023_Waidele_Zhao} and over $6\,\leq\,m\,\leq\,10$ by \citet{2024_Lekshmi_Gizon_Jain_etal} who reported stronger power amplitudes during solar maximum.
Additionally, we detect that the cross power decreases with increasing depth from the near-surface reference depth down to approximately $8.5-9.7$\,Mm.
This decrease varies $\approx\,6-12.5\%$ for the TD and RD data, respectively.
This is qualitatively consistent with the decrease in Fourier power found by \citet{2020_Proxauf_Gizon_Loptien_etal} (i.e., $\approx$\,10$\%$) down to 8\,Mm \added{and down to 11\,Mm by \citet{2018_Loptien_Gizon_Birch_etal}, but not in agreement with work by \citet{2024_Mandal_Hanasoge}.}

Next, we discuss the normalized phases between the near-surface reference depth and subsequent depths.
Our results show that in both the long-term average phase behavior (Fig.\,\ref{fig:RD_longtermphase} and Fig.\,\ref{fig:TD_longtermphase}) as well as in the temporal phase evolution (Fig.\,\ref{fig:phase_difference_m} and Fig.\,\ref{fig:TD_phase_difference_m}) for select $m$, two consistent trends emerge.
First, we find that deeper depths lead in phase over shallower ones (i.e., negative $\Delta\varphi_m(t, d_{0i})/m$).
Second, we observe a systematic pattern of increasing negative phase with depth.
Taken together, this is consistent with Rossby waves exhibiting a retrograde tilt relative to the Sun's rotation that is constant with depth.
We note that this negative phase shift indicates a longitudinal or horizontal displacement of the wave structure with depth, so it provides information regarding the structural tilting of Rossby waves, rather than energy propagation. 

However, we find no strong and consistent evidence for the modulation of $\Delta\varphi_m(t, d_{0i})/m$ by the solar cycle.
While this suggests a largely stable tilted wave structure across the solar cycle, longer-term observations are necessary to confirm any definitive connection between the phase behavior and solar activity.
Notably, prior work by \citet{2024_Lekshmi_Gizon_Jain_etal} reported that the correlation between Rossby wave power and solar activity was weaker during SC23 than SC24, highlighting the need for additional cycle coverage.

Our results also extend those of \citet{2020_Proxauf_Gizon_Loptien_etal}, who examined phase variations in seven years of RD data down to 16\,Mm below the surface from the temporal Fourier transform of the spherical harmonic coefficients of radial vorticity.
The authors reported that the phase of the radial eigenfunctions for all $m$ remain nearly constant with depth ($\pm\,5\,\degree$), while background phases vary strongly, showing depth-dependent changes up to $\pm\,100\,\degree$.
While we compare our results qualitatively to those of \citet{2020_Proxauf_Gizon_Loptien_etal}, as theirs represent the only other available estimate of Rossby wave phase differences in the literature, we want to emphasize \added{an important methodological distinction.}
\added{In their analysis, the temporal Fourier phases reflect both the structural tilt of the mode and its temporal evolution, as the Fourier transform inherently mixes these contributions.}
\added{In contrast, we compute phase differences directly from the spherical harmonic coefficients associated with the coherent Rossby wave signal, yielding phases that reflect only the depth-dependent tilt of the modes, without the temporal variability incorporated by the Fourier transform.}
\added{{Additionally, we note that \citet{2020_Proxauf_Gizon_Loptien_etal} evaluated the phase differences using a single frequency bin, whereas our analysis includes multiple frequency bins, which may reduce noise and make subtle depth-dependent tilt trends more apparent.}}

Consistent with their results, we also report small phases (around $-0.01$\,rad to $-0.02$\,rad), particularly for $6\,\leq\,m\,\leq\,11$.
However, in contrast to \citeauthor{2020_Proxauf_Gizon_Loptien_etal}'s findings of constant depth-dependent behavior for all $m$, our analysis reveals distinct depth-dependent variations in $\Delta\varphi_m(t,d_{0i})/m$, with the magnitude of these variations depending on $m$.
The fact that some $m$ show stronger modulation than others further suggests a mode dependence, where larger-scale modes may couple more efficiently to the global magnetic field or are more affected by rotation rates.
The sensitivity of different $m$ to the solar cycle seen in the computation of frequency shifts caused by the Sun's internal rotation rate was shown in \citet{2020_Goddard_Birch_Fournier_Gizon}.

We also note that, to maintain an acceptable temporal cadence, SHD must be performed on a partially observed Sun, which is a common practice in global helioseismology. 
However, this approach is known to introduce power leakage, which may affect the phase measurements central to this study. 
To assess the influence of such leakage on the phase differences we measure between Rossby waves at different depths, we construct a toy model. 
In this model, we numerically generate two sets of artificial Rossby waves that follow the dispersion relations and power distributions corresponding to two depths, as illustrated in Fig.\,\ref{fig:RD_Fourier_Maps}. 
Phase differences, defined as functions of azimuthal order $m$, are inserted between the two depth-dependent wave sets in the simulated time series. 
We then restrict the data to a partial solar disk of the same size as used in our main analysis and apply the identical analysis procedure. 
The results demonstrate that the procedure successfully recovers the inserted phase differences with a precision better than $0.1\%$ across all $m$. 
It is also found that the recovery precision improves with increasing $m$. 
We therefore conclude that the power leakage associated with partially observed Sun analysis does not significantly affect the measured phase differences.

Beyond this, our results can be understood within the broader framework of Rossby wave properties and characteristics.
While classical (hydrodynamical) equatorial Rossby waves are shown to have a largely columnar or vertical structure \citep[e.g.,][]{1982_Saio, 2019_Liang-Gizon_Birch_Duvall}, previous studies have reported small but significantly nonzero imaginary components in their radial eigenfunctions \citep[e.g.,][]{2020_Proxauf_Gizon_Loptien_etal, 2020_Gizon_Fournier_Albekioni}, indicating subtle phase shifts with depth.
Additionally, when magnetic fields are included in simulations, longitudinal phase shifts with height emerge for magnetized Rossby waves \citep{2017_Dikpati_Cally_McIntosh_Heifetz, 2018_Dikpati_McIntosh_Bothun_etal, 2024_Dikpati_Gilman_Raphaldini_McIntosh}.
In particular, the introduction of a longitudinal magnetic field induces a tilt in the Rossby wave structure, breaking their otherwise columnar structure \citep{2018_Dikpati_McIntosh_Bothun_etal, 2024_Dikpati_Gilman_Raphaldini_McIntosh}.
As illustrated in Fig.\,2 of \citet{2017_Dikpati_Cally_McIntosh_Heifetz} and Fig.\,6 of \citet{2018_Dikpati_McIntosh_Bothun_etal}, the velocity and magnetic perturbations from the mean differential rotation and magnetic fields exhibit longitudinal phase offsets, which enable angular momentum transport via Reynolds or Maxwell stresses in opposite directions.
These stress imbalances drive nonlinear energy exchanges between the mean differential rotation, magnetic fields, and Rossby waves, manifesting as tilted, oscillatory Rossby wave patterns.
This suggests that the retrograde tilt with depth observed in our near-surface Rossby waves results from their coupling with the differential rotation in the near-surface shear layer and/or large-scale magnetic fields, but further work is required to quantify the extent of that interaction.

Observationally, \citet{2022_Harris_Dikpati_Hewins_etal} and \citet{2024_Raphaldini_Dikpati_McIntosh_Teruya} also report coronal structures associated with magneto-Rossby waves drifting in longitude independently of differential rotation. 
\citeauthor{2022_Harris_Dikpati_Hewins_etal} identified retrograde tilts in long-lived coronal holes linked to Rossby modes, which is qualitatively consistent with our results.
Comparable behavior, including small tilts and inferred vertical energy propagation of Rossby modes, has also been noted in Earth's atmosphere \citep[e.g.,][]{1991_Ebisuzaki, 2012_Holton_Hakim, 2017_Vallis}, as well as simulations of Titan \citep[e.g.,][]{2023_Lewis_Lombard_Read_etal}, Venus \citep[e.g.,][]{2006_Imamura, 2019_Kashimura_Sugimoto_Takagi_Matsuda_etal}, and general terrestrial atmospheric models \citep[e.g.,][]{2016_DiasPinto_Mitchell}.
This tilt has implications for angular momentum transport, similarly to the solar context.

From a physical standpoint, the picture that emerges is that Rossby waves are tilted retrograde to the Sun's rotation, with deeper layers leading in phase over shallower ones.
The longitudinal phase offset increases systematically with depth, with Rossby waves appearing to maintain a stable inclined structure throughout the solar cycle.
This depth-dependent phase suggests that Rossby waves are valuable diagnostic tools for probing the Sun's interior dynamics, including angular momentum and energy transport processes.
Further work that combines observational constraints from near-surface depths with extended MHD simulations, such as those of \citet{2024_Dikpati_Gilman_Raphaldini_McIntosh}, may help quantify the extent to which coupling with the magnetic field and/or differential rotation contributes to the observed retrograde tilt of Rossby waves.

\begin{acknowledgments}
We thank Dr.~Matthias Waidele for providing the codes used to prepare the helioseismic flow field data analyzed in this study. The HMI data used in this publication are courtesy of NASA's SDO and the HMI science team. HMI is an instrument developed by Stanford University under the NASA contract number NAS5-02139. 
This work is funded by NASA DRIVE Science Center COFFIES project under grant number 80NSSC22M0162.
\end{acknowledgments}

\facilities{SDO(HMI)}

% \software{astropy \citep{2013A&A...558A..33A,2018AJ....156..123A,2022ApJ...935..167A},  
%           Cloudy \citep{2013RMxAA..49..137F}, 
%           Source Extractor \citep{1996A&AS..117..393B}
%           }


\appendix
\restartappendixnumbering
\section{Time-Distance Results} \label{sec:time_distance_appendix}

The corresponding TD results are presented below.

\subsection{Fourier Power, Cross Power, and Coherence Spectra} \label{subsec:TD_Fourier_Maps}

\begin{figure*}[ht!]
    \centering
\includegraphics[width=\linewidth]{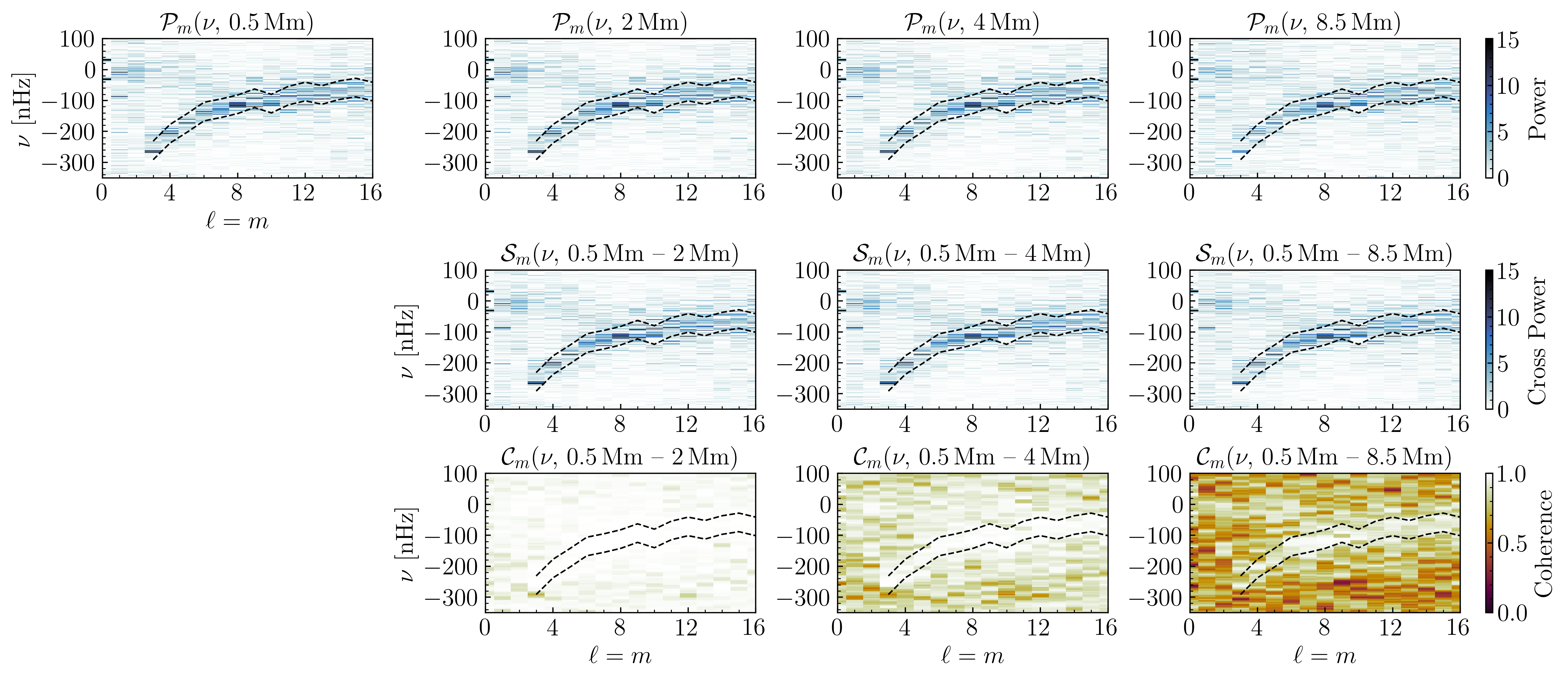}
    \caption{Sectoral power $\mathcal{P}_m(\nu, d)$, cross power $\mathcal{S}_m(\nu, d_{0i})$, and coherence $\mathcal{C}_{m}(\nu, d_{0i})$ spectra of the decomposed TD radial vorticity data, focusing on azimuthal orders $3\,\leq\,m\,\leq\,16$. The TD data is tracked at the Carrington rotation rate. The figure is the same as Fig.\,\ref{fig:RD_Fourier_Maps}.\label{fig:TD_Fourier_Maps}}
\end{figure*}

The temporal Fourier power, cross power, and coherence spectra for the TD data are shown in Fig.\,\ref{fig:TD_Fourier_Maps}. 
The results are consistent with the RD data in Fig.\,\ref{fig:RD_Fourier_Maps}.

\subsection{Long-term Average Depth-Dependent Phases} \label{subsec:TD_longterm_phases}

\begin{figure*}[ht!]
    \centering
        \includegraphics[scale=0.5]{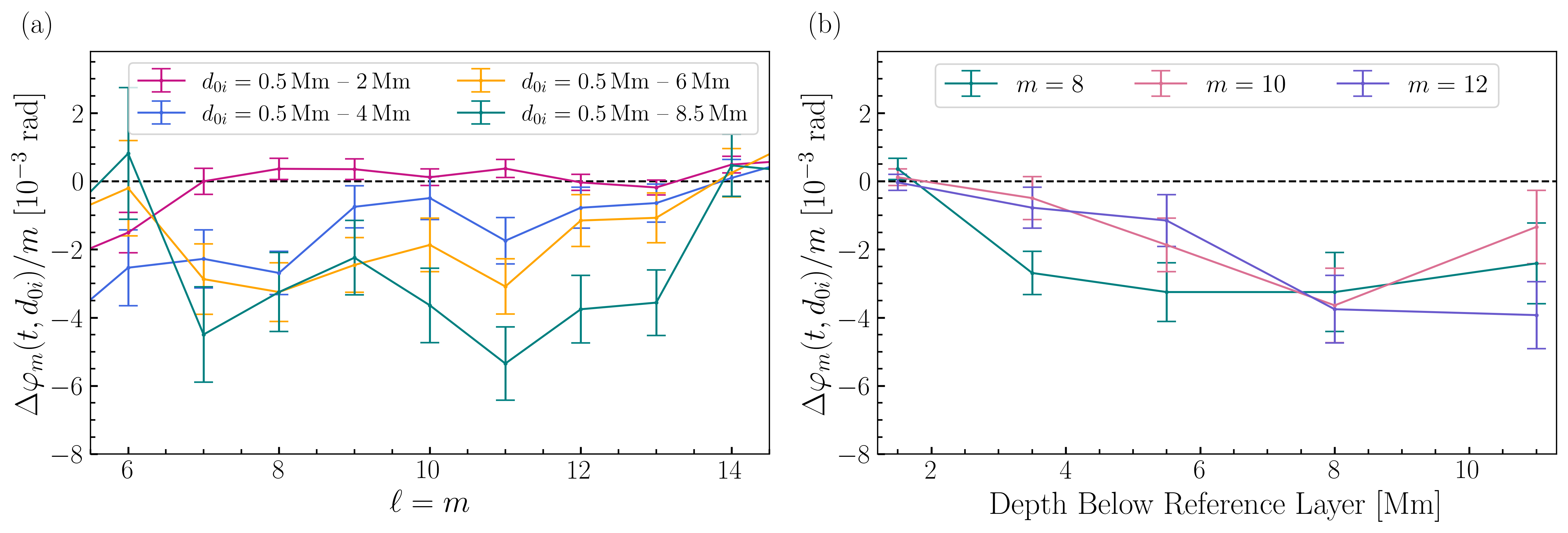}
  \caption{Normalized phase differences $\Delta\varphi_m(t, d_{0i})/m$ derived from the filtered complex coefficients of the decomposed TD radial vorticity between the near-surface reference depth (0.5\,Mm) and several subsurface depths, averaged over approximately 14.5\,yr. (a): $\Delta\varphi_m(t, d_{0i})/m$ versus $\ell=m$, focusing on $6\,\leq\,m\,\leq\,14$. (b): $\Delta\varphi_m(t, d_{0i})/m$ versus depth below the near-surface reference depth for $m = 8$, $10$, and $12$. Error bars represent the standard error across time for $\Delta\varphi_m(t, d_{0i})/m$ at each depth and $m$.}
    \label{fig:TD_longtermphase}
\end{figure*}

The long-term averaged depth-dependent $\Delta\varphi_m(t, d_{0i})/m$ for the TD data is presented in Fig.\,\ref{fig:TD_longtermphase}.

\subsection{Temporal Evolution of Depth-Dependent Cross Power and Phase}

\begin{figure*}[ht!]
\centering
        \includegraphics[scale=0.5]{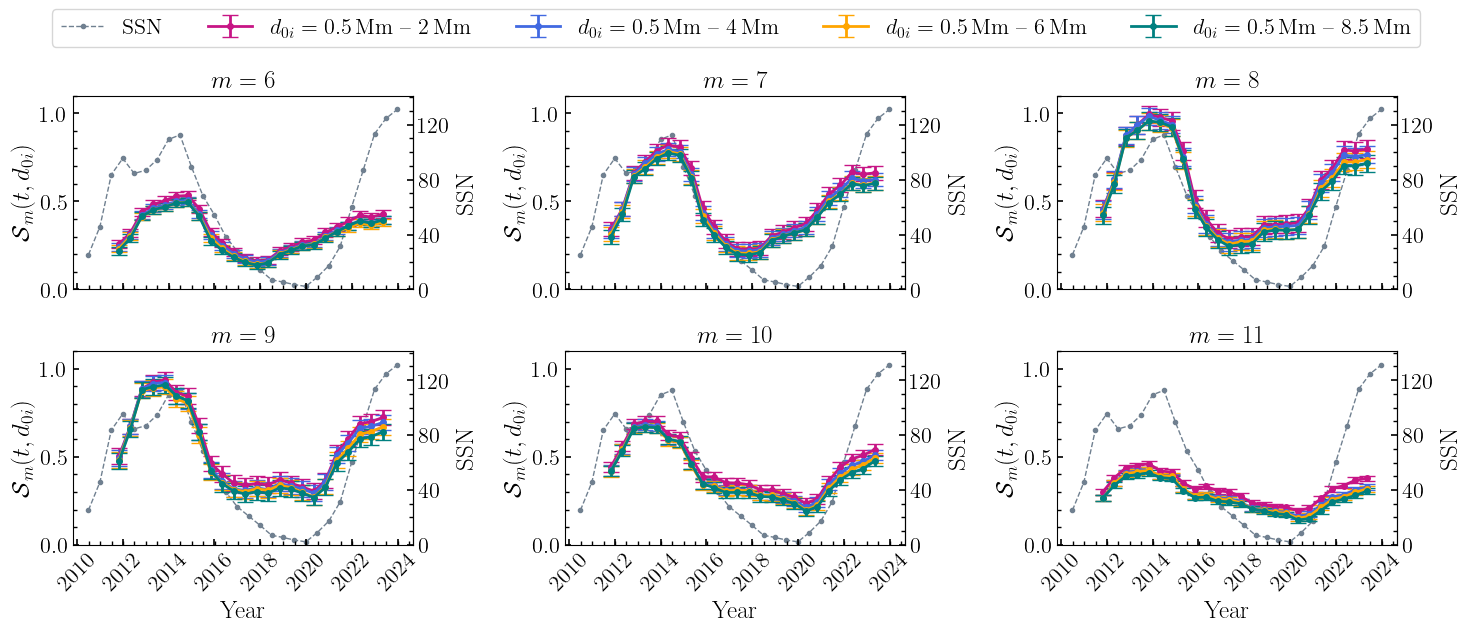}
\caption{Temporal evolution of the cross power $\mathcal{S}_m(t, d_{0i})$ for $6\,\leq\,m\,\leq\,11$ between the near-surface reference depth (0.5\,Mm) and select subsurface depths for the TD data. For all depths, the dataset is segmented into 3\,yr moving windows with a 0.5\,yr step size. Error bars represent standard errors across time segments. The SSN (gray) is overplotted for comparison with the solar cycle.}
\label{fig:TD_cross_power_temporal}
\end{figure*}

\begin{figure*}[ht!]
    \centering
    \includegraphics[scale=0.5]{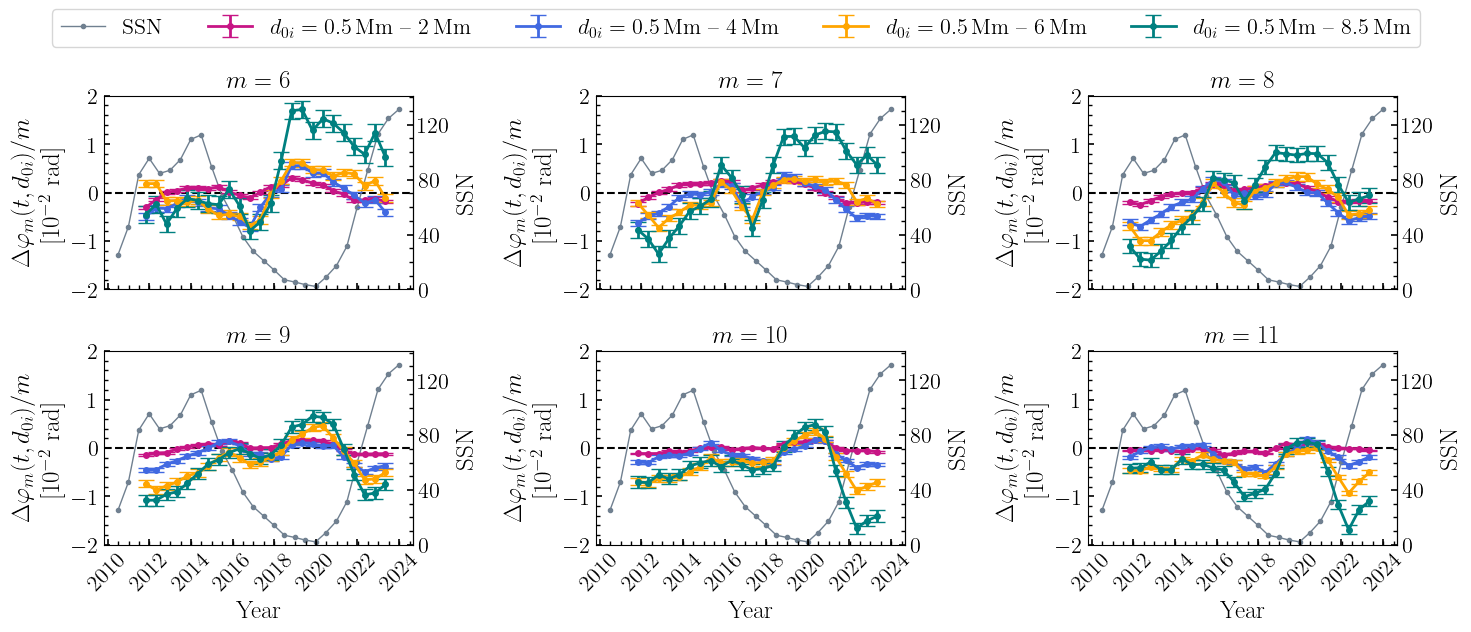}
    \caption{Temporal evolution of the normalized phase differences $\Delta\varphi_m(t, d_{0i})/m$ for $6\,\leq\,m\,\leq\,11$ between the near-surface reference depth (0.5\,Mm) and select subsurface depths for the TD data. For all depths, the dataset is segmented into 3\,yr moving windows with a 0.5\,yr step size. Error bars represent standard errors across time segments. The SSN (gray) is overplotted for comparison with the solar cycle.}
    \label{fig:TD_phase_difference_m}
\end{figure*}

The temporal evolution of the cross power computed from the TD data is presented in Fig.\,\ref{fig:TD_cross_power_temporal} and Fig.\,\ref{fig:TD_phase_difference_m}.

%% For this sample we use BibTeX plus aasjournalv7.bst to generate the

\clearpage
% \newpage

\bibliography{refs}{}
\bibliographystyle{aasjournalv7}



\end{document}